\documentclass[aps,pra,amsmath,amssymb,reprint]{revtex4-1}

\usepackage{graphics}
\usepackage{dcolumn,bm}

\begin{document}

\title{Comment on ``Bose-Einstein condensation with a finite number of particles in a power-law trap''}

\author{Jos\'{e} M. B. Noronha}
\email{jnoronha@por.ulusiada.pt}
\affiliation{Universidade Lus\'{\i}ada - Porto, Rua Dr. Lopo de Carvalho, 4369-006 Porto, Portugal}

\date{November 26, 2014}

\begin{abstract}
In Jaouadi \textit{et al.} [Phys. Rev. A \textbf{83}, 023616 (2011)] the authors derive an analytical finite-size expansion for the Bose-Einstein condensation critical temperature of an ideal Bose gas in a generic power-law trap. In the case of a harmonic trap, this expansion adds higher order terms to the well-known first order correction. We point out a delicate point in connection to these results, showing that the claims of Jaouadi \textit{et al.} should be treated with caution. In particular, for a harmonic trap, the given expansion yields results that, depending on what is considered to be the critical temperature of the finite system, do not generally improve on the established first order correction. For some non-harmonic traps, the results differ at first order from other results in the literature.

\end{abstract}

\pacs{03.75.Hh, 05.30.Jp}

\maketitle

The critical temperature for Bose-Einstein condensation (BEC) of an ideal gas trapped in a generic cartesian power-law potential in the thermodynamic limit was given in \cite{Bagnato1987}. In the case of power-law traps with spherical or axial symmetry, this temperature was given in \cite{Salasnich2000,Jaouadi2010}. We will denote it by $T_c^0$ (the superscript 0 meaning thermodynamic limit). Following the first realizations of BEC with dilute trapped gases, corrections to $T_c^0$ due to the finite number of trapped particles were given, in the case of an isotropic harmonic trap, in \cite{GrossmannHolthaus1995b,*GrossmannHolthaus1995c}. Shortly after, this was generalized to the anisotropic harmonic trap in \cite{KetterlevanDruten1996}, the result reading
\begin{equation}
\frac{\Delta T_c}{T_c^0}=-\frac{\zeta(2)}{2\zeta(3)^{2/3}}\frac{\omega_a}{\omega_g}N^{-1/3}\; ,
\label{classicalcorrection}
\end{equation}
where $\zeta$ is the Riemann zeta function, $\omega_a$ and $\omega_g$ are the arithmetic and geometric means of the trap frequencies respectively and $N$ is the number of particles.

Actually, it is well known that for such a system (with a finite number of particles) BEC does not strictly exist as a sharp, mathematically defined phase transition. Instead, it gradually happens over a narrow temperature window as the temperature is lowered. It follows that eq.~(\ref{classicalcorrection}) cannot be taken too seriously. It merely provides a reference value for the location of this temperature window (and it is very good at doing that).

In \cite{Jaouadi2011} the authors report an analytical finite-size expansion for $\Delta T_c/T_c^0$, which is given formally to all orders, in powers of $x_0\equiv E_0/(k_BT_c)$, where $E_0$ is the single particle ground level energy. Their framework is that of a general anisotropic power-law trap. 
In the specific case of a harmonic trap, we have $x_0=(3\omega_a)/(2\omega_g)(\zeta(3)/N)^{1/3}(T_c^0/T_c)$ and truncating at the second correction term, their result reads
\begin{equation}
\frac{\Delta T_c}{T_c^0}= -\frac{\zeta(2)}{2\zeta(3)^{2/3}}\frac{\omega_a}{\omega_g}N^{-1/3}
+\zeta\left(\frac{3}{2}\right)\left(\frac{2\omega_a^3}{3\pi\zeta(3)\omega_g^3}\right)^{1/2}N^{-1/2}
\; .
\label{Jaouadicorrection}
\end{equation}
This is eq.~(30) in \cite{Jaouadi2011} (apart from a small difference in notation for $\omega_a$ and $\omega_g$). We see that this results extends the older one to higher order. The authors note that the second
term in (\ref{Jaouadicorrection}) is of order $N^{-1/2}$ instead of the more expectable $N^{-2/3}$, which makes the new correction important for $N\lesssim 10^5$.
They also present specific results for some non-harmonic power-law traps. In this comment, we point out that these results depend on how the critical temperature is defined for the finite system. Specifically, other possible approaches can lead to conclusions that differ significantly from the conclusions of \cite{Jaouadi2011}, for some choices of the parameters.

In order to obtain numerical values for the BEC critical temperature of a finite system, some specific definition for this temperature, which works for such systems, is necessarily involved. Two common criteria in the literature are
the maximum of the specific heat curve against temperature (vanishing first derivative of $C(T)$) and the inflection point of the condensate fraction curve against temperature (vanishing second derivative of $N_{\text{gr}}/N(T)$). We will denote the critical temperatures obtained using these two criteria by $T_{\textrm{max}}$ and $T_{\textrm{infl}}$, respectively. A multitude of possible criteria and their appropriateness, including the two above, was discussed early on in \cite{PajkowskiPathria1977}, the authors concluding in favor of the specific heat maximum as the best one. After the first successful BEC experiments with trapped gases, one of those authors \cite{Pathria1998} considered again the criteria for BEC, this time in the specific context of harmonic traps, concluding in favor of eq.~(\ref{classicalcorrection}) as is (without further correction terms), as providing the best definition of a critical temperature. Nevertheless, we stress that this is a subjective matter.

The criterium used in \cite{Jaouadi2011} is implicit in the adopted procedure. The authors work in the context of the local density approximation. In this setting, the thermodynamic sums over states are approximated by integrals in phase space, which are then integrated over the momentum, yielding thermodynamic quantities in the form of spatial integrals. The critical temperature was then obtained by setting the critical value of the chemical potential at $\mu_c=E_0$ and omitting from the spatial integral for $N$ the trap region where the potential is less than $E_0$ (the thermodynamic limit critical temperature is recovered by setting $E_0=0$). The physical significance of this criterium for the critical temperature is unclear to us (as is does not seem to be based on a specific physical attribute of the system) and motivated the numerical investigation of different possible criteria which we now undertake.

\begin{figure}
\includegraphics{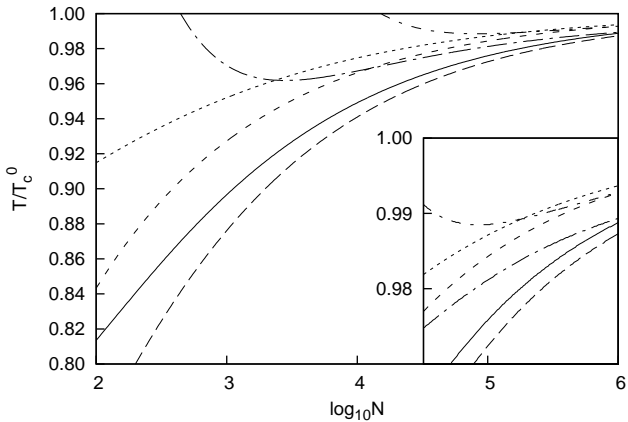}
\caption{\label{fig:isotropic}Rescaled temperature $T_{\textrm{max}}/T_c^0$ (solid line), $T_{\textrm{infl}}/T_c^0$ (long-dashed line), $T_c/T_c^0$ according to eq.~(\ref{classicalcorrection}) (short-dashed line), $T_c/T_c^0$ according to \cite{Jaouadi2011}
(dotted line), $T_{1\%}/T_c^0$ (long-dashed/dotted line) and $T_{0.1\%}/T_c^0$ (short-dashed/dotted line). See text for definitions of these temperatures. The inset zooms in on the large $N$ region. It shares a common horizontal axis with the larger plot.}
\end{figure}

We have computed numerically, directly from the thermodynamic sums, $T_{\textrm{max}}$ and $T_{\textrm{infl}}$. We also computed two additional temperatures, which might serve as reference points: the temperatures at which the condensate fraction is $1\%$ and $0.1\%$, which we denote by $T_{1\%}$ and $T_{0.1\%}$ respectively. In Fig.~\ref{fig:isotropic}
we present the ratio of these four temperatures to $T_c^0$ as functions of the number of particles for an isotropic harmonic trap. Also shown are the usual first order correction, given by eq.~(\ref{classicalcorrection}), and the higher order correction of \cite{Jaouadi2011}, given to order $N^{-1/2}$ in eq.~(\ref{Jaouadicorrection}). For the numerics, we have actually used also the three next order terms of this expansion (not shown in eq.~(\ref{Jaouadicorrection})) to assure its convergence.
We see that the two criteria, specific heat maximum and inflection point of condensate fraction, give similar
results, with $T_c/T_c^0$ lowering as the system becomes smaller. As for the temperatures at fixed 
condensate fraction, $T_{1\%}$ and $T_{0.1\%}$, 
as the system becomes smaller they inflect upwards at some point. This is 
a reflection of the fact 
that as the system becomes smaller the phase transition becomes increasingly diluted and the condensate 
fraction eventually becomes significant even above $T_c^0$. 
The corrections to $T_c^0$ given by eq.~(\ref
{classicalcorrection}) and by the result of \cite{Jaouadi2011} follow intermediate paths. We see that the ambiguity in what can be considered as the transition region is very marked towards the low $N$ range. In particular, except perhaps for large $N$, it does not seem very meaningful to talk about \textit{the} critical temperature (unless we previously fix it with some -- necessarily arbitrary -- definition). We must point out that the authors in \cite{Jaouadi2011} did not consider values of $N$ lower than $10^3$ (their Fig.~1 has the $N$ range $10^3$--$10^6$).
 
\begin{figure}
\includegraphics{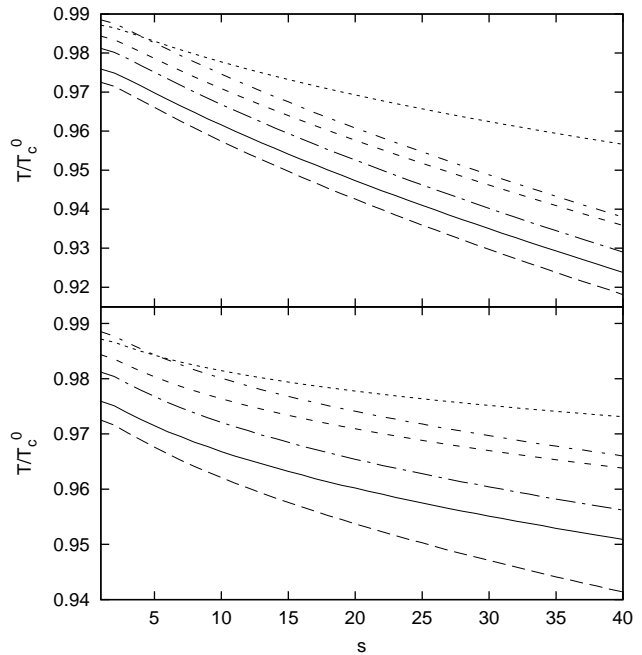}
\caption{\label{fig:anisotropic}The same rescaled temperatures as in Fig.~\ref{fig:isotropic} are shown 
as functions of the anisotropy parameter $s$ for the case of $N=10^5$ particles in a disc shaped (upper plot) and cigar shaped (lower plot) harmonic trap. The meaning of the several curve patterns is the same as in Fig.~\ref{fig:isotropic}.}
\end{figure}

We now consider anisotropic traps.
In Fig.~\ref{fig:anisotropic}
we plot the same quantities for axially symmetric disc shaped and cigar shaped harmonic traps, with anisotropy factor $s\equiv \omega_i/\omega_j$ (where $i=\textrm{axial}$, $j=\textrm{radial}$ for the disc trap and the opposite for the cigar trap), as functions of $s$ for $N=10^5$. We see that as the anisotropy increases the transition happens at increasingly lower $T/T_c^0$. However, unlike when $N$ is lowered, when $s$ is increased the $T_{1\%}$ and $T_{0.1\%}$ curves do not deviate significantly from the $T_{\textrm{max}}$ and $T_{\textrm{infl}}$ curves. This is because in this case $T_c/T_c^0$  is reduced \textit{without} diluting the phase transition, which remains sharp. 
We note that for the highly anisotropic cigar shaped trap, the critical temperature given by the result of  \cite{Jaouadi2011} deviates clearly from the critical temperatures given by the other criteria we have considered, whereas the one given by eq.~(\ref{classicalcorrection}) does not.
(If the expansion of \cite{Jaouadi2011} were truncated at order $N^{-1/2}$, i.e., eq.~(\ref{Jaouadicorrection}), the respective curve would be slightly above its current position in both plots of Fig.~\ref{fig:anisotropic}, therefore deviating slightly more from the other curves.)

Summarizing, in the case of a harmonic trap, the result for $\Delta T_c/T_c^0$ given by \cite{Jaouadi2011} does not seem to provide a better pointer to the transition region than the first order result of eq.~(\ref{classicalcorrection}). However, due to the lack of a single, well-defined critical temperature, this observation is somewhat subjective (except in the case of the highly anisotropic disc shaped trap, where the result of \cite{Jaouadi2011} falls above the transition region). Overall, we believe it is not helpful to talk about higher order finite-size correction terms to the critical temperature, unless we previously fix it by adopting some physical criterium for its definition.
Likewise, the claim in \cite{Jaouadi2011} that eq.~(\ref{classicalcorrection}) for the critical temperature is innacurate for $N\leq 10^5$, the new term in eq.~(\ref{Jaouadicorrection}) becoming significant in this range, is not meaningful on its own, without the respective criterium.

In what regards non-harmonic power-law traps, we limit ourselves to pointing out the existence of other results in the literature which are in disagreement with the result of \cite{Jaouadi2011} for power-law potentials between cubic and power 6 already at first order.
Specifically, in \cite{KirstenToms1998} the first order result $\Delta T_c/T_c^0=-x_0\zeta(\eta)/[(\eta +1)\zeta(\eta +1)]$, where we are using the notation of \cite{Jaouadi2011}, was derived  
for general power-law potentials satisfying $\eta>1$. $\eta$ is a constant defined in \cite{Jaouadi2011} that is related only to the powers of the power-law trap. $\eta>1$ roughly means a confining potential less steep than a power-law of power 6. Eq.~(\ref{classicalcorrection}) is the $\eta=2$ particular case of the above formula. The authors in \cite{KirstenToms1998} used a high temperature finite-size expansion for $N$ which takes into 
account the discreteness of the energy levels. The simpler 
method of \cite{PethickSmith} yields exactly the same result. Although in \cite{PethickSmith} the end result was presented for the harmonic potential only, the generalization to power-law potentials with $\eta>1$ is trivial from the intermediate results therein. The method of \cite{PethickSmith} was used in \cite{Gautam2008} in the specific case of a quartic power-law trap (for which $\eta=5/4$), yielding the $\eta=5/4$ instance of the same formula. Now, this formula clashes with the result of \cite{Jaouadi2011} in the cases where $1<\eta \leq 3/2$ (potential between cubic and power 6). Indeed, while the first order shift $\Delta T_c/T_c^0$ it predicts is linear in $x_0$ and negative, the one predicted by \cite{Jaouadi2011} for $\eta <3/2$ is of order $x_0^{\eta-1/2}$, therefore more pronounced, and positive. Agreement at first order 
happens only for $\eta>3/2$ (roughly, power-laws less steep than cubic, including the harmonic trap).

The disagreement for $1<\eta \leq 3/2$ might be understandable on the grounds that the critical temperature is definition dependent (or ill-defined) for finite systems. However, note that this means that the conclusions drawn in \cite{Jaouadi2011} for non-harmonic traps also depend on the critical temperature definition.
In particular, from their expansion the authors conclude that in the $\eta=3/2$ case (e.g., a cubic power-law such as $V(r)=ar^3$) the finite-size effects on $T_c$ completely cancel and that for very steep power-laws the critical temperature shift can be very significant and positive, with possible implications for the experimental realization of BEC in such traps. These conclusions are contrary to the finite-size corrections obtained in \cite{KirstenToms1998,PethickSmith,Gautam2008}. Thus, they cannot be taken in a strict sense and should be considered with caution. 
Moreover, the method of \cite{KirstenToms1998} seems to us more adequate, in principle, for finite-size calculations than the method of \cite{Jaouadi2011}, since it takes into account the discreteness of the energy levels.
We believe this particular matter deserves further investigation.


%

\end{document}